\documentclass[aps,prl,twocolumn,superscriptaddress]{revtex4-1}

\usepackage{dcolumn}
\usepackage{amsmath,amssymb,amsthm,paralist}
\usepackage{graphicx}
\usepackage{appendix}
\usepackage{subfigure}
\usepackage{longtable}
\usepackage{url}
\usepackage{verbatim}

\begin{document}


\title{Mass measurements demonstrate a strong $N=28$
shell gap in argon}


\author{Z.~Meisel}
\email[]{meisel@nscl.msu.edu}
\affiliation{National Superconducting Cyclotron Laboratory, Michigan State University, East Lansing, Michigan, USA}
\affiliation{Department of Physics and Astronomy, Michigan State University, East Lansing, Michigan, USA}
\affiliation{Joint Institute for Nuclear Astrophysics, Michigan State University, East Lansing, Michigan, USA}
\author{S.~George}
\affiliation{National Superconducting Cyclotron Laboratory, Michigan State University, East Lansing, Michigan, USA}
\affiliation{Joint Institute for Nuclear Astrophysics, Michigan State University, East Lansing, Michigan, USA}
\affiliation{Max-Planck-Institut f\"{u}r Kernphysik, Heidelberg, Germany}
\author{S.~Ahn}
\affiliation{National Superconducting Cyclotron Laboratory, Michigan State University, East Lansing, Michigan, USA}
\affiliation{Joint Institute for Nuclear Astrophysics, Michigan State University, East Lansing, Michigan, USA}
\author{J.~Browne}
\affiliation{National Superconducting Cyclotron Laboratory, Michigan State University, East Lansing, Michigan, USA}
\affiliation{Department of Physics and Astronomy, Michigan State University, East Lansing, Michigan, USA}
\affiliation{Joint Institute for Nuclear Astrophysics, Michigan State University, East Lansing, Michigan, USA}
\author{D.~Bazin}
\affiliation{National Superconducting Cyclotron Laboratory, Michigan State University, East Lansing, Michigan, USA}
\author{B.A.~Brown}
\affiliation{National Superconducting Cyclotron Laboratory, Michigan State University, East Lansing, Michigan, USA}
\affiliation{Department of Physics and Astronomy, Michigan State University, East Lansing, Michigan, USA}
\author{J.F.~Carpino}
\affiliation{Department of Physics, Western Michigan University,
Kalamazoo, Michigan, USA}
\author{H.~Chung}
\affiliation{Department of Physics, Western Michigan University,
Kalamazoo, Michigan, USA}
\author{R.H.~Cyburt}
\affiliation{National Superconducting Cyclotron Laboratory, Michigan State University, East Lansing, Michigan, USA}
\affiliation{Joint Institute for Nuclear Astrophysics, Michigan State University, East Lansing, Michigan, USA}
\author{A.~Estrad\'{e}}
\affiliation{School of Physics and Astronomy, The University of Edinburgh, Edinburgh, UK}
\author{M.~Famiano}
\affiliation{Department of Physics, Western Michigan University,
Kalamazoo, Michigan, USA}
\author{A.~Gade}
\affiliation{National Superconducting Cyclotron Laboratory, Michigan State University, East Lansing, Michigan, USA}
\affiliation{Department of Physics and Astronomy, Michigan State University, East Lansing, Michigan, USA}
\author{C.~Langer}
\affiliation{National Superconducting Cyclotron Laboratory, Michigan State University, East Lansing, Michigan, USA}
\affiliation{Joint Institute for Nuclear Astrophysics, Michigan State University, East Lansing, Michigan, USA}
\author{M.~Mato\v{s}}
\altaffiliation[Present address: ]{Physics Division, International Atomic Energy Agency, Vienna, Austria}
\affiliation{Department of Physics and Astronomy, Louisiana State University, Baton Rouge, Louisiana, USA}
\author{W.~Mittig}
\affiliation{National Superconducting Cyclotron Laboratory, Michigan State University, East Lansing, Michigan, USA}
\affiliation{Department of Physics and Astronomy, Michigan State University, East Lansing, Michigan, USA}
\author{F.~Montes}
\affiliation{National Superconducting Cyclotron Laboratory, Michigan State University, East Lansing, Michigan, USA}
\affiliation{Joint Institute for Nuclear Astrophysics, Michigan State University, East Lansing, Michigan, USA}
\author{D.J.~Morrissey}
\affiliation{National Superconducting Cyclotron Laboratory, Michigan State University, East Lansing, Michigan, USA}
\affiliation{Department of Chemistry, Michigan State University, East Lansing, Michigan, USA}
\author{J.~Pereira}
\affiliation{National Superconducting Cyclotron Laboratory, Michigan State University, East Lansing, Michigan, USA}
\affiliation{Joint Institute for Nuclear Astrophysics, Michigan State University, East Lansing, Michigan, USA}
\author{H.~Schatz}
\affiliation{National Superconducting Cyclotron Laboratory, Michigan State University, East Lansing, Michigan, USA}
\affiliation{Department of Physics and Astronomy, Michigan State University, East Lansing, Michigan, USA}
\affiliation{Joint Institute for Nuclear Astrophysics, Michigan State University, East Lansing, Michigan, USA}
\author{J.~Schatz}
\noaffiliation{}
\author{M.~Scott}
\affiliation{National Superconducting Cyclotron Laboratory, Michigan State University, East Lansing, Michigan, USA}
\affiliation{Department of Physics and Astronomy, Michigan State University, East Lansing, Michigan, USA}
\author{D.~Shapira}
\affiliation{Oak Ridge National Laboratory, Oak Ridge, Tennessee, USA}
\author{K.~Smith}
\altaffiliation[Present address: ]{Department of Physics and Astronomy, University of Tennessee, Knoxville, Tennessee, USA}
\affiliation{Joint Institute for Nuclear Astrophysics, Michigan State University, East Lansing, Michigan, USA}
\affiliation{Department of Physics, University of Notre Dame, South Bend, Indiana, USA}
\author{J.~Stevens}
\affiliation{National Superconducting Cyclotron Laboratory, Michigan State University, East Lansing, Michigan, USA}
\affiliation{Department of Physics and Astronomy, Michigan State University, East Lansing, Michigan, USA}
\affiliation{Joint Institute for Nuclear Astrophysics, Michigan State University, East Lansing, Michigan, USA}
\author{W.~Tan}
\affiliation{Department of Physics, University of Notre Dame, South Bend, Indiana, USA}
\author{O.~Tarasov}
\affiliation{National Superconducting Cyclotron Laboratory, Michigan State University, East Lansing, Michigan, USA}
\author{S.~Towers}
\affiliation{Department of Physics, Western Michigan University, Kalamazoo, Michigan, USA}
\author{K.~Wimmer}
\altaffiliation[Present address: ]{Department of Physics, University of Tokyo, Tokyo, Japan}
\affiliation{National Superconducting Cyclotron Laboratory, Michigan State University, East Lansing, Michigan, USA}
\author{J.R.~Winkelbauer}
\affiliation{National Superconducting Cyclotron Laboratory, Michigan State University, East Lansing, Michigan, USA}
\affiliation{Department of Physics and Astronomy, Michigan State University, East Lansing, Michigan, USA}
\author{J.~Yurkon}
\affiliation{National Superconducting Cyclotron Laboratory, Michigan State University, East Lansing, Michigan, USA}
\author{R.G.T.~Zegers}
\affiliation{National Superconducting Cyclotron Laboratory, Michigan State University, East Lansing, Michigan, USA}
\affiliation{Department of Physics and Astronomy, Michigan State University, East Lansing, Michigan, USA}
\affiliation{Joint Institute for Nuclear Astrophysics, Michigan State University, East Lansing, Michigan, USA}


\date{\today}

\begin{abstract}

We present results from recent time-of-flight nuclear mass
measurements at the National Superconducting
Cyclotron Laboratory at Michigan State University.
We report the first mass measurements of $^{48}$Ar and
$^{49}$Ar and find atomic mass excesses of -22.28(31)~MeV and
-17.8(1.1)~MeV, respectively. These masses provide strong evidence
for the closed shell nature of neutron number $N=28$ in argon, which is therefore the lowest
even-$Z$ element exhibiting the $N=28$ closed shell. The 
resulting trend in binding-energy differences, which probes the
strength of the $N=28$ shell, compares favorably with shell-model calculations in
the sd-pf shell using SDPF-U and SDPF-MU Hamiltonians.

\end{abstract}


\maketitle


The ``magic" numbers of protons and neutrons, which enhance nuclear binding for
isotopes near the valley of $\beta$-stability, can evolve for more neutron-rich or neutron-deficient
nuclei~\cite{Brow01,Sorl08,Caki13}.
The neutron magic number $N=28$ has been the subject of extensive
recent experimental and theoretical
investigations~\cite{Sorl13,Nowa09,Gaud10,Utsu12,Caur14}.
Since neutron-rich $N=28$ nuclei are within
experimental reach and are computationally tractable for shell-model
calculations, they are ideal candidates for illuminating the
fundamental forces at work in exotic nuclei.
It is known that the $N=28$ shell gap, which stabilizes
doubly magic $^{48}_{20}$Ca$^{}_{28}$, is absent in the $Z=14$ and
$Z=16$
isotopic chains at
$^{42}_{14}$Si$^{}_{28}$~\cite{Bast07,Camp06,Take12,Stro14} and
$^{44}_{16}$S$^{}_{28}$~\cite{Glas97,Sara00,Gaud09,Forc10,Sant11}. 
Experimental information on the structure of $^{40}_{12}$Mg$^{}_{28}$ 
suggests it has a prolate deformed ground state~\cite{Craw14}, which
would be consistent with the absence of a neutron shell gap.

The existence of the
$N=28$ shell gap for argon is a matter of some controversy.
Several previous experimental studies have assessed the
shell structure of neutron-rich 
argon~\cite{Sche96,Gade03,Grev03,Gade05,Gaud06,Gaud08,Bhat08,Gade09,Meng10,Wink12,Cali14}.
Investigation of the energy of the lowest excited states of
$^{45}_{18}$Ar$^{}_{27}$
via $\beta$-decay spectroscopy of
$^{45}_{17}$Cl$^{}_{28}$ suggested 
a weakened, but still present, $N=28$ shell closure for
argon~\cite{Grev03}.
The first $2^{+}$
state energies $E(2_{1}^{+})$ along the argon isotopic
chain~\cite{Rama01,Bhat08,Gade09} and
information on neutron single-particle structure from
transfer~\cite{Gaud06,Gaud08} and knockout~\cite{Gade05} reactions
are consistent with the presence of an $N=28$ shell gap in
$^{46}_{18}$Ar$^{}_{28}$. 
 Though disagreement exists as
to the inferred nuclear structure from measurements of the
$^{46}_{18}$Ar$^{}_{28}$ quadrupole
excitation strength, $B(E2,0_{1}^{+}\rightarrow2_{1}^{+})$, written
as $B(E2)$ hereafter for brevity. 
Three projectile Coulomb excitation measurements, two at
intermediate energies~\cite{Sche96,Gade03} and one at
Coulomb-barrier beam energy~\cite{Cali14}, deduce a low $B(E2)$,
corresponding to a reduced quadrupole collectivity. In this case
quadrupole collectivity reflects a propensity for
neutrons to be excited across the $N=28$ shell gap, and thus a
low $B(E2)$ may be expected for a semi-magic nucleus.
State-of-the-art shell-model calculations that properly account for
the breakdown of the $N=28$ magic number in silicon and sulfur
isotopes predict a markedly higher $B(E2)$ for
$^{46}$Ar~\cite{Wink12}. A low-statistics lifetime measurement of
the $2_{1}^{+}$ state of $^{46}$Ar deduced a high $B(E2)$ value in
agreement with theory~\cite{Meng10}, but at odds with the three
consistent, independent Coulomb excitation
measurements~\cite{Sche96,Gade03,Cali14}.

However, $B(E2)$ measurements are not necessarily unambiguous probes
of neutron shell structure, since they are sensitive to proton
degrees of freedom and proton-neutron interactions.  In contrast,
mass measurements, and the neutron separation energies derived from
them, directly probe the neutron shell gap in a model-independent
way.

We report here results from the first~\footnote{The review of shell structure in the
$N=28$ region \cite{Sorl13} lists the mass of $^{48}_{18}$Ar$^{}_{30}$
as measured, though the link they cite pointing to an AME version
from 2011 is no longer in
operation.  The more recent 2012 Atomic Mass Evaluation~\cite{Audi12}
lists the mass of $^{48}_{18}$Ar$^{}_{30}$ as unmeasured and we are
unable to find experimental data for this mass after a detailed
search of the literature.}
mass measurements of $^{48}$Ar and $^{49}$Ar, which provide robust
evidence for the persistence of the $N=28$ shell gap for argon.
These results
were obtained with the time-of-flight
(TOF) technique at the National Superconducting Cyclotron Laboratory
(NSCL)~\cite{Mato12,Estr11,Meis13}.
Neutron-rich isotopes of silicon to zinc
were produced by fragmentation of a 140~MeV/u $^{82}$Se
primary beam impinging on a beryllium target. A target thickness of 517~mg/cm$^{2}$ was
used to produce less neutron-rich nuclei, required for calibration,
whereas a target thickness of 658~mg/cm$^{2}$ was used to produce the
more neutron-rich fragments of interest.  The fragments were
transmitted through the A1900 fragment separator~\cite{Morr03} to
the focal plane of the S800 spectrograph~\cite{Bazi03}. A
7.5~mg/cm$^{2}$ Kapton wedge degrader was used in the A1900 to
remove the high flux of low-$Z$ nuclei that would otherwise complicate fragment
identification. The thick
and thin targets were used alternately, while the magnetic rigidity
$B\rho$ of the A1900 beam-line and the S800 were left unchanged.  This
allowed us to measure the TOF for nuclei with a broader range of
mass-to-charge ratios $m/q$.
By design, the lower $m/q$ isotopes observed generally had well-known masses and could
be used to calibrate the relationship between $m/q$ and TOF, whereas
the higher $m/q$ nuclei observed generally had unknown masses. TOF was measured
over a 60.6~m flight path using fast timing scintillators located at the A1900 and S800 focal
planes. A typical TOF was $\approx500$~ns. The finite momentum spread of the beam, limited to $\delta p/p=\pm$0.5\% by
slits in the A1900, made a precise measurement of $B\rho=p/q$ necessary
for each nucleus produced.  $B\rho$ was measured by detecting the
position of each ion at a dispersive focus at the S800 target
position. Position measurements were performed by collecting
electrons emitted from a gold foil due to passing beam particles on
a position sensitive microchannel plate detector~\cite{Shap00}.
The energy loss measurement obtained from the ionization chamber in
the S800 focal plane combined with TOF provided fragment identification.
 
 \begin{figure}[ht]
 \includegraphics[width=1.0\columnwidth,angle=0]{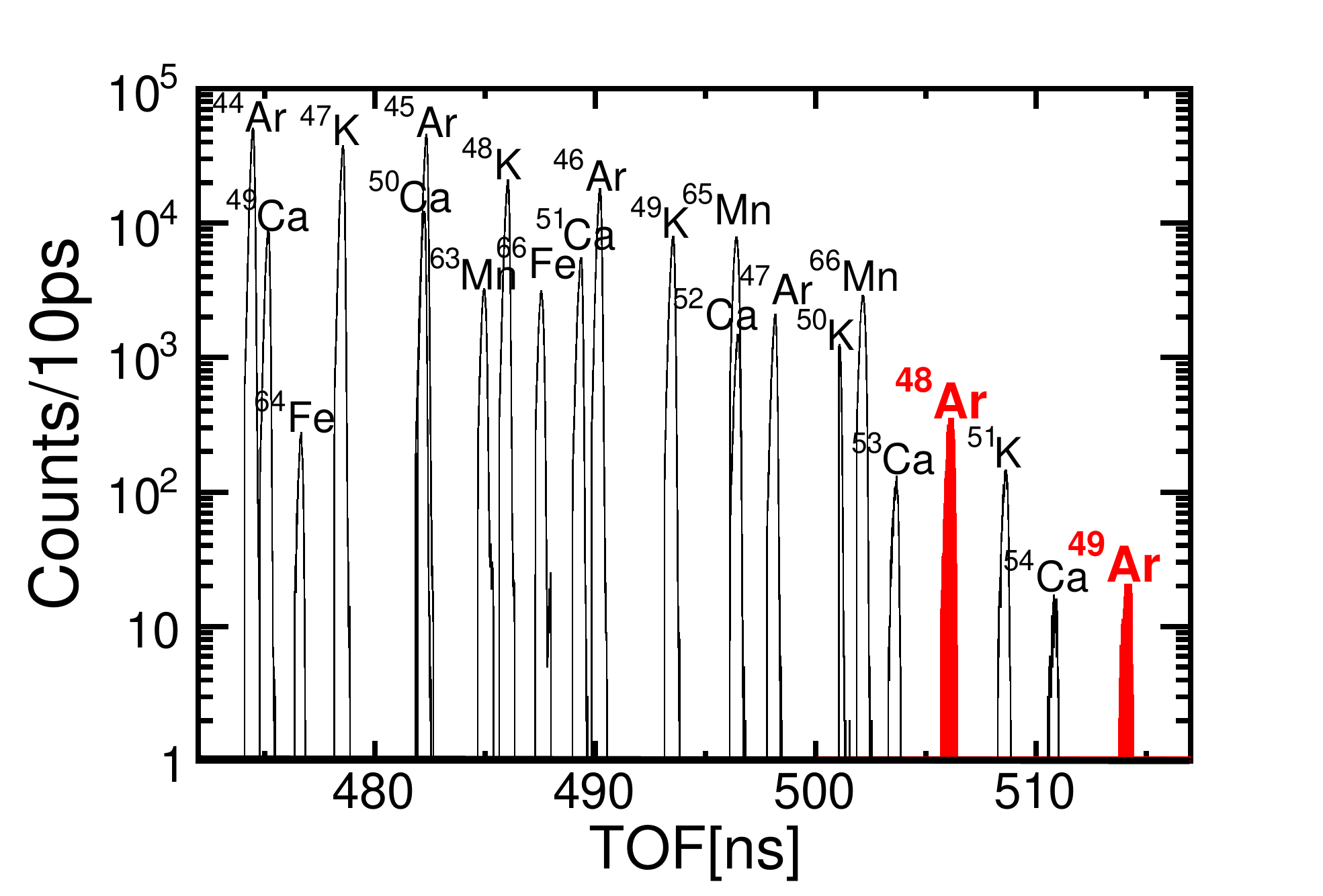}
 \caption{(color online). Rigidity-corrected time-of-flight distributions for
 reference nuclei (unfilled histograms) used to calibrate the $\frac{m_{rest}}{q}(TOF)$
 relationship to obtain masses from TOFs of $^{48}$Ar and $^{49}$Ar
 (red-filled histograms).
 \label{TOFspectrum}}
 \end{figure}

 \begin{figure}[ht]
 \includegraphics[width=1.0\columnwidth,angle=0]{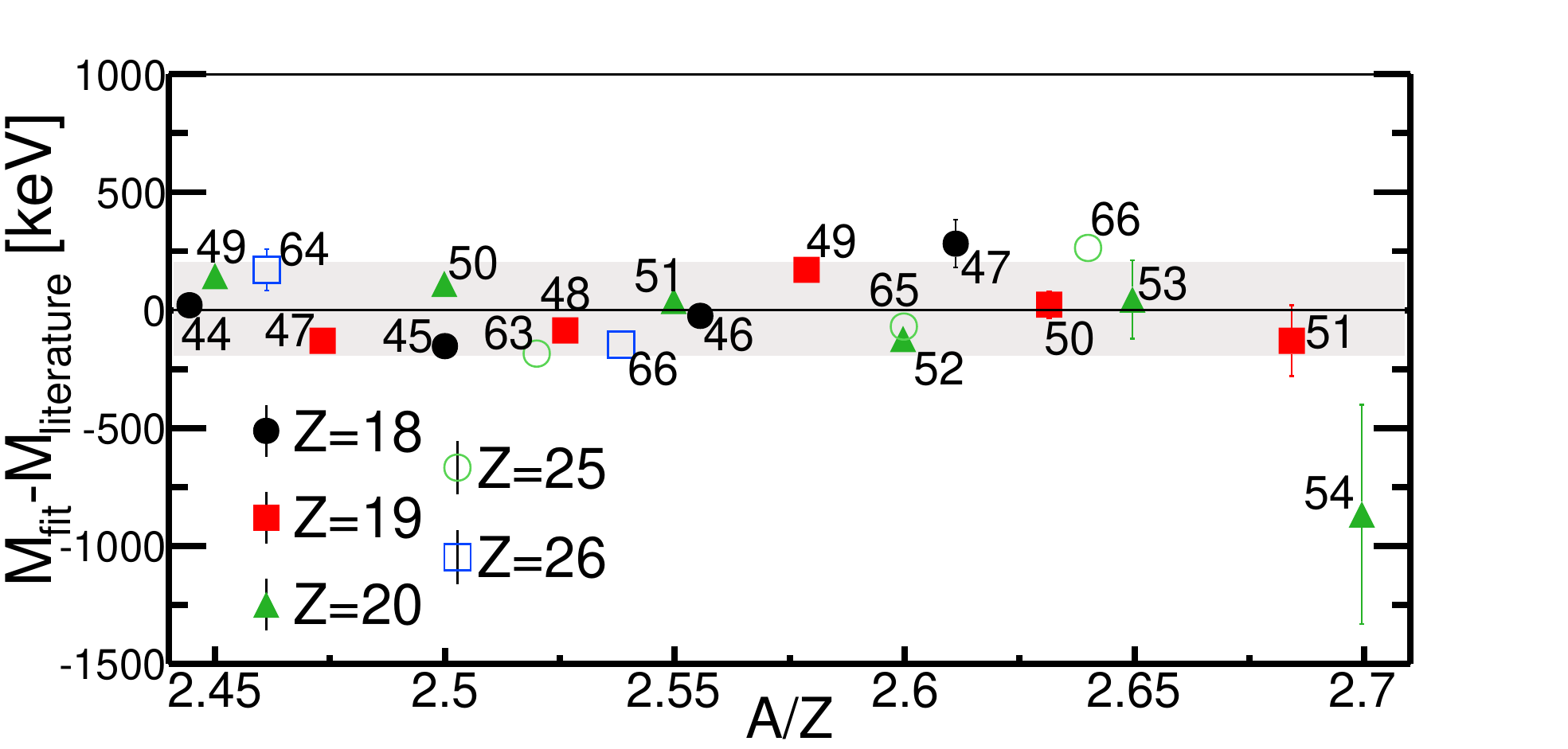}
 \caption{(color online). Residuals of the fit to the time-of-flight of calibration
 nuclei (discussed further in the text) as a function of the
 mass number to nuclear charge ratio $A/Z$. Isotopes are labeled
 with their mass
 number and symbols indicate the elements (solid circle
 for argon,
 solid square for potassium, solid triangle for calcium, open circle
 for manganese, and
 open square for iron).
 Calibration masses were fit to within 9~keV/q without any
 systematic trends. The gray band shows the average systematic
 mass uncertainty included for reference nuclei as described in
 ~\cite{Mato12}.\label{MassFit}}
 \end{figure}

In principle, the simultaneous measurement of an ion's TOF, charge $q$, and $B\rho$
through a magnetic system of a known flight path $L_{path}$ directly
yields its mass,
$m_{rest}=\frac{TOF}{L_{path}}\frac{q(B\rho)}{\gamma}$, where
$\gamma$ is the Lorentz factor. 
However, in practice neither $L_{path}$ nor the ion optical
dispersion used to determine $B\rho$ are known with sufficient
precision. Furthermore, only a measurement of $B\rho$
relative to the central ion optical axis is performed. Therefore,
the $\frac{m_{rest}}{q}(TOF)$ relationship is determined empirically
using reference nuclei with well-known masses~\cite{Meis13}.
The TOF distributions for reference nuclei and $^{48,49}$Ar are shown
in Figure~\ref{TOFspectrum}.
Twenty reference nuclei with masses known
to better than 100~keV and no known isomeric states longer lived than
100~ns~\cite{Audi12,Wein13,Audi12B} were fitted with a 7-parameter calibration function of
second order in TOF, first order in TOF*$Z$, and containing first,
second, and fourth order $Z$ terms.
The calibration function represents a minimal set of terms that
minimized the overall fit residual
to literature masses and resulted in no detectable systematic
biases~\cite{Mato12}, as seen in Figure~\ref{MassFit}. Note that the
apparently deviant point $^{54}$Ca in Figure~\ref{MassFit} does not
significantly impact the results of the mass fit, due to its large
statistical uncertainty. A systematic uncertainty
of 9.0~keV/q was included as described in~\cite{Mato12} to normalize
the
$\chi^{2}$ per degree of freedom of the mass fit to one.
Two additional uncertainties related to the extrapolation were added
to the final mass uncertainties, one to reflect the uncertainties in the TOFs of
reference nuclei, which leads to an uncertainty in the fit
coefficients of the $\frac{m_{rest}}{q}(TOF)$ relation, and one to reflect
the uncertainty inherent in
choosing a particular calibration function over another which has a
comparable goodness of fit.
The latter was determined by investigating the robustness of the
results to adding additional terms to the calibration function.
The total mass uncertainty is a sum in quadrature of
statistical, systematic, and two extrapolation uncertainties. The
relative contribution of the extrapolation uncertainties becomes
larger as the distance in $m/q$ and $Z$ from reference nuclei
increases.


The atomic mass excesses obtained for $^{48}$Ar and $^{49}$Ar were
-22.28(31)~MeV and -17.8(1.1)~MeV, respectively.
This corresponds to a measurement precision of $\delta m/m\approx~10^{-5}$.
These masses can now be used as a probe of shell structure~\cite{Lunn03}.
Typically, binding-energy differences of neutron-rich nuclei are examined 
for this purpose in order to
isolate the impact of adding neutrons.
One such probe that is frequently used is the two-neutron separation
energy $S_{2n}$.
$S_{2n}(Z,A)=2*ME_{neutron}+ME(Z,A-2)-ME(Z,A)$, where $ME$ is the
mass excess, represents the energy required to remove two neutrons
from a nucleus with $Z$ protons and $A-Z$ neutrons. Along an
isotopic chain, $S_{2n}$
generally declines with increasing $N$ due to the liquid-drop aspect
of nuclear binding that penalizes a large neutron-proton asymmetry.
This decline is markedly increased following a nucleus that
exhibits a magic neutron number. 
However, the change in slope that indicates a shell closure is not
always easy to interpret. A recently introduced
quantity $D_{n}$~\cite{Brow13},
where $D_{n}(Z,A)=(-1)^{N+1}[S_{n}(Z,A+1)-S_{n}(Z,A)]$, provides a more
readily recognizable signature of a shell closure. In a given mass
region, $D_{n}$ indicates
the number of orbital angular momentum projection ``$m$" states that
participate in pairing for a given nucleus.  A peak in $D_{n}$
at a certain neutron number along with a change in the $D_{n}$ level before
and after that neutron number indicates a shell gap~\cite{Brow13}.
The change in the $D_{n}$ level is a crucial element since it
indicates a transition from filling one ``$m$" state to filling another.
 

 
 \begin{figure}
 \includegraphics[width=0.95\columnwidth,angle=0]{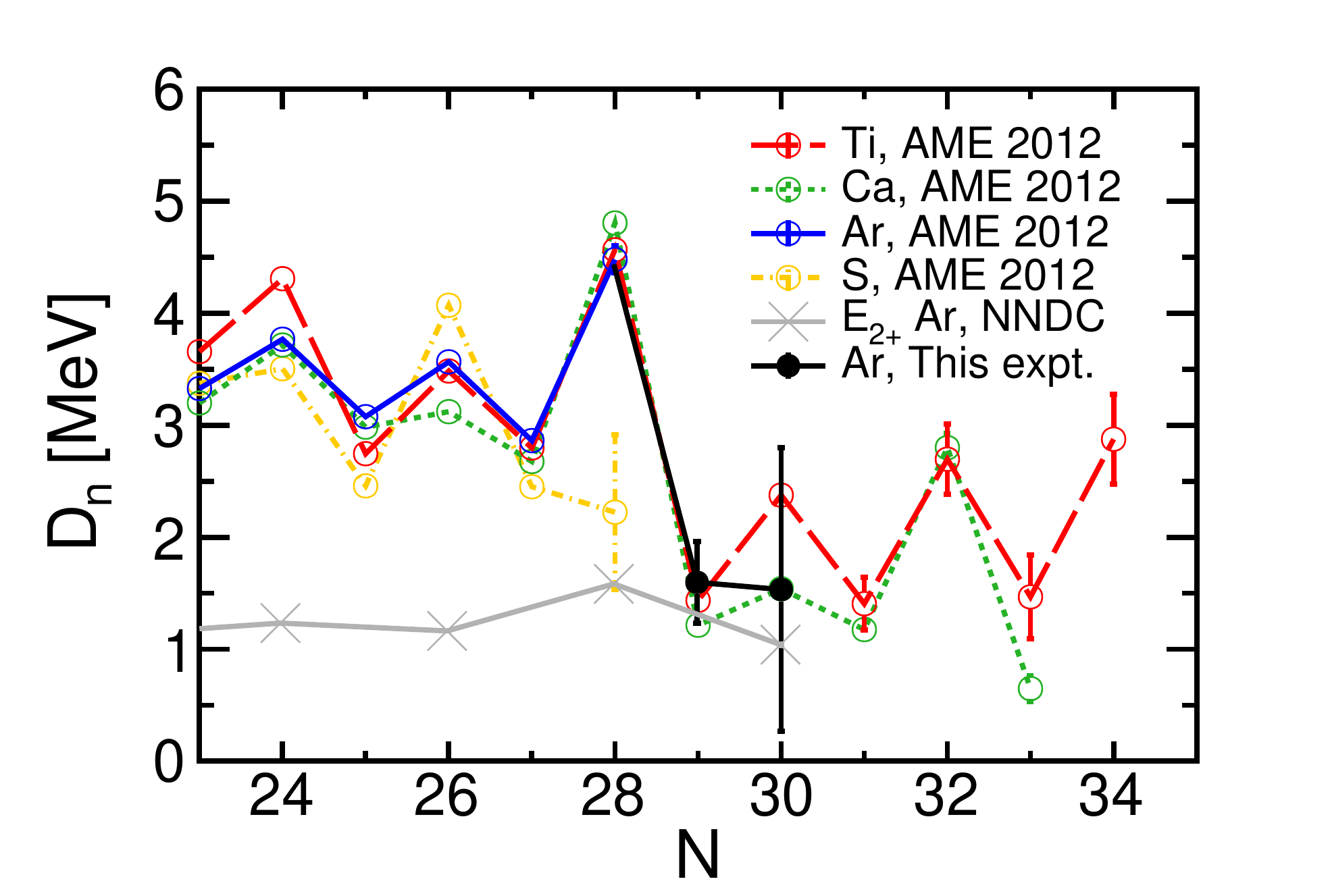}
 \caption{(color online). $D_{n}$~\cite{Brow13} as a function of
 neutron number $N$
 near $N=28$ for sulfur (dot-dash
 line), argon (solid
 line), calcium (dotted line), and
 titanium (dashed line).  The previously known~\cite{Audi12} argon trend
 (solid line, open circles) is shown along with results from this
 experiment (solid line, solid circles).  $E(2_{1}^{+})$
 energies~\cite{Rama01,Bhat08,Gade09} are shown for
 comparison (crosses).  The peak at $N=28$ followed by a reduction in
 $D_{n}$ for $N>28$ as compared to $N<28$ indicates the presence of a closed
 shell. From shell-model calculations we conclude the transition
 from $D_{n}\approx3$~MeV for $N<28$ to $D_{n}\approx1.5$~MeV for
 $N>28$ corresponds to the
 transition from filling the $f_{7/2}$ orbit to filling the $p_{3/2}$
 orbit.\label{ArDnTrend}}
 \end{figure}

The $D_{n}$ values for argon isotopes from this work show a clear
signature for an $N=28$ shell closure (Figure~\ref{ArDnTrend}).
With the new mass
excesses for $^{48,49}$Ar, it is apparent that neutron-rich argon displays the same
systematics in $D_{n}$ as calcium and titanium, which are known
to exhibit an $N=28$ shell gap~\cite{Sorl13}. As seen in Figure~\ref{ArDnTrend},
sulfur does not peak at $N=28$~\cite{Sara00}, which is consistent with prior
conclusions that sulfur does not exhibit the $N=28$ closed
shell~\cite{Glas97}. Based on our experimental data we can therefore
conclude that argon
is the lowest even-$Z$ element with a closed neutron shell for $N=28$.

\begin{figure}
 \includegraphics[width=0.95\columnwidth,angle=0]{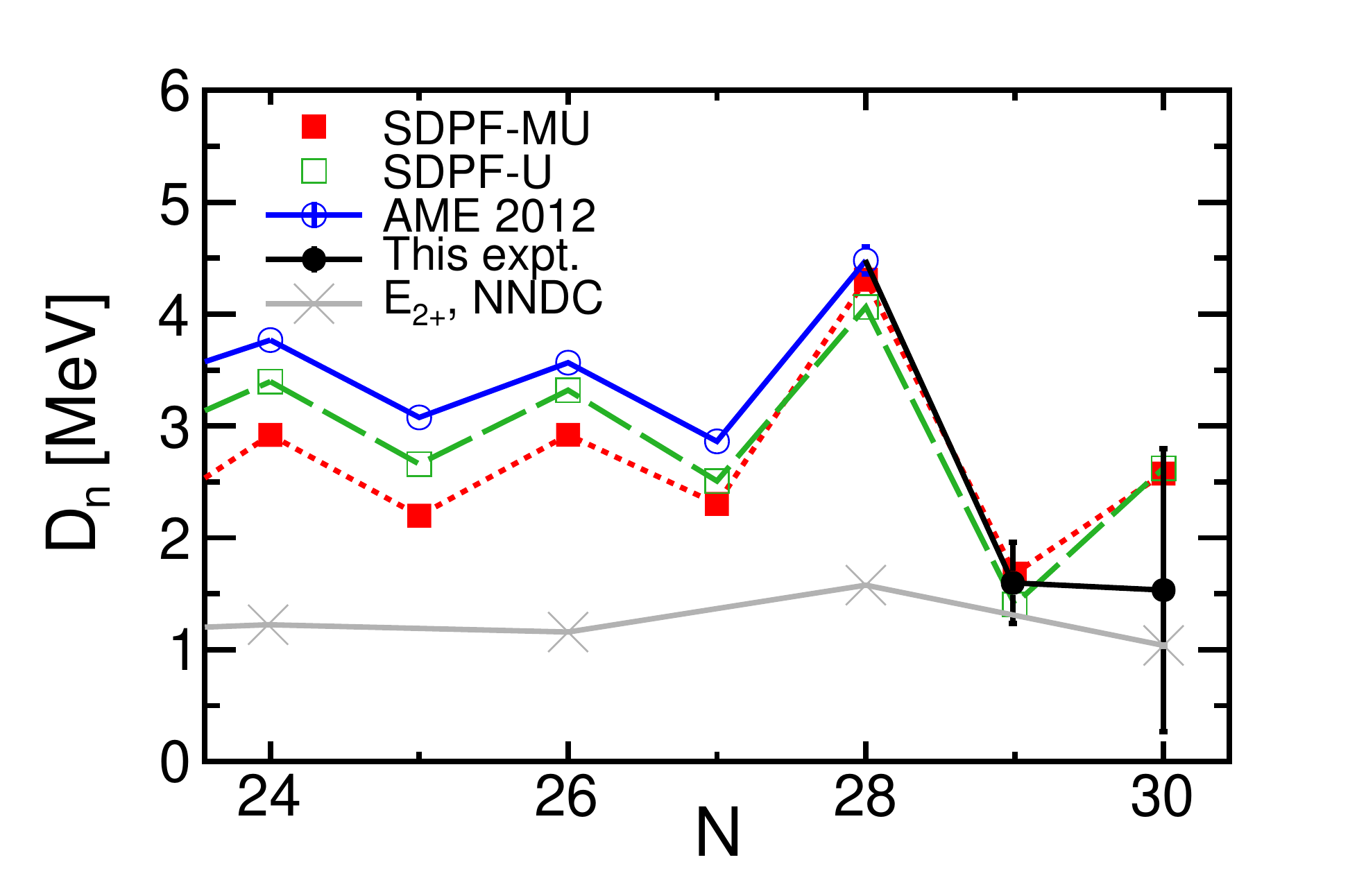}
 \caption{(color online). The $D_{n}$~\cite{Brow13} trend near $N=28$ for argon from
 currently known masses~\cite{Audi12} (open circles) and this
 experiment (solid circles) is shown along with shell-model
 calculations employing the SPDF-MU Hamiltonian~\cite{Utsu12}
 (solid squares) and
 the SDPF-U Hamiltonian~\cite{Nowa09} (open squares).
 $E(2_{1}^{+})$ energies~\cite{Rama01,Bhat08,Gade09} are shown for
 comparison (crosses).
 \label{ArDnShellModel}}
 \end{figure}

We compare the experimental $D_{n}$ trend to the $D_{n}$ trends
for local mass predictions obtained from shell-model calculations using the
SDPF-U~\cite{Nowa09} and SDPF-MU~\cite{Utsu12} interactions in
Figure~\ref{ArDnShellModel}. In both
cases there is excellent agreement between experiment and theory.
This indicates current shell-model calculations adequately describe
the interaction between core and valence neutrons around $N=28$ for
argon. 

In summary, we performed the first mass measurements of $^{48}$Ar
and $^{49}$Ar via the time-of-flight technique. We find the $N=28$
closed shell is present for argon, which makes argon the lowest
even-$Z$ element that exhibits an $N=28$ shell gap.
Based on this result we can conclude that the problems of
shell model calculations in describing electromagnetic observables
in argon isotopes near $N=28$ are not related to the neutron shell
gap, but instead points to issues with the interaction of valence
neutrons and core protons.

\begin{acknowledgments}
This project is based upon work supported by the National Science
Foundation under Grant No. PHY-0822648, PHY-1102511, PHY-1404442, and
PHY-1430152. S.G. acknowledges support from the DFG under Contract
No.
GE2183/1-1 and GE2183/2-1.
\end{acknowledgments}

\bibliographystyle{apsrev4-1}
\bibliography{TOFmassArStructureMeisel}


\end{document}